\documentclass[reprint]{revtex4-1}

\usepackage{hyperref,amsmath,amssymb,siunitx,color,graphicx}

\begin{document}

\title{High speed determination of laser wavelength using Poincar\'e descriptors of speckle}

\author{Laura O'Donnell}
\author{Kishan Dholakia}
\author{Graham D. Bruce}
\email{gdb2@st-andrews.ac.uk}

\affiliation{SUPA School of Physics and Astronomy, University of St Andrews, North Haugh, St Andrews, UK, KY16 9SS}
\date{\today}
\begin{abstract}
Laser speckle can provide a powerful tool that may be used for metrology, for example measurements of the incident laser wavelength with a resolution beyond that which may be achieved in a commercial device. However, to realise highest resolution requires advanced multi-variate analysis techniques, which limit the acquisition rate of such a wavemeter. Here we show an arithmetically simple method to measure wavelength changes with dynamic speckle, based on a Poincar\`e descriptor of the speckle pattern. We demonstrate the measurement of wavelength changes at femtometer-level with a measurement time reduced by two orders of magnitude compared to the previous state-of-the-art, which offers promise for applications such as speckle-based laser wavelength stabilisation. 
\end{abstract}

\maketitle


\section{Introduction}

Shaping of the spatial profile of light to a pre-determined pattern of interest has allowed major advances in diverse areas such as optical manipulation, imaging and communications \cite{RubinszteinDunlop16}. A commonly underappreciated method of beam shaping is to randomize the spatial profile of the light to create a speckle pattern. The randomized intensity pattern has been extensively used in the field of ultracold atoms as a controlled source of noise or disorder \cite{Billy08,Choi16,Thomson16}. Moreover, as the interference processes leading to the formation of speckle are linear, deterministic and reversible \cite{Dainty13}, dynamic changes in the speckle pattern can be exploited as an exquisite sensor for changes in the scattering medium \cite{Kaufmann11,Boas10,Briers13}, or in the illuminating light field itself \cite{Kohlgraf10,Mazilu12,Mourka13,Mazilu14,Metzger17,Cao17,Alexeev17,Bruce19,Meng19}.

Here, we explore a new method to use dynamic speckle to probe changes of the wavelength of a laser beam. The formation of speckle can be thought of as a highly complex interferometer, where the speckle grains are the analog of interferometer fringes. These grains thus have a wavelength-dependent size and position and can therefore, after an appropriate training procedure, be used as a marker for wavelength. The transmission matrix method \cite{Cao17,Meng19} is capable of measuring these changes over a range as large as \SI{400}{\nano\meter} to \SI{1100}{\nano\meter} at a resolution down to the speckle correlation limit, which can be as low as \SI{1}{\pico\meter}, while principal component analysis (PCA) \cite{Mazilu14,Metzger17,Bruce19} can be used to improve this resolution down to the attometer level. This high-resolution and broad range compares well with current high-end wavemeters, while offering a simple, compact and relatively inexpensive setup. This configuration also realises a speckle-based spectrometer with picometer-level resolution \cite{Cao17}.

Among the application areas for such a high-accuracy wavemeter is the option of wavelength-stabilising a laser without the need for a spectroscopic reference \cite{Couturier18}. Stabilising the wavelength of a laser using a speckle-based setup locks only to a speckle pattern of interest. Therefore, the lock-point is freely chosen. However, the maximum lock update rate previously achieved using PCA is \SI{200}{\hertz} \cite{Metzger17}. This is relatively slow compared to modern laser frequency controllers which are designed to provide current feedback to compensate \SI{}{\kilo\hertz}-rate frequency noise, necessitating the search for a faster speckle analysis technique.

A change in wavelength on a scale below the speckle correlation limit causes only grain-scale changes to the speckle pattern. Recently, a technique involving Poincar\'e descriptors was developed to investigate the size of speckle grains \cite{Majumdar17, Majumdar18}. Here we show that the grain-scale variations of speckle patterns that are induced by changes of the wavelength $\lambda$ are also captured by a Poincar\'e descriptor. Below the correlation limit, this metric varies monotonically with wavelength, making it a suitable tool for measuring wavelength in this regime. We demonstrate the measurement of femtometer-level wavelength changes using laser speckle generated from an integrating sphere. Finally, we show that the method offers a huge increase in computation speed relative to PCA, making it an interesting candidate for devices which require accurate, \SI{}{\kilo\hertz}-rate measurements of wavelength.

\section{Poincar\'e descriptor for wavelength measurement}

To carry out Poincar\'e analysis, each speckle pattern $I_{x,y}\left(\lambda\right)$, where $x,y$ denote spatial co-ordinates, is normalised with respect to its maximum value in order to remove intensity fluctuations between patterns. The Poincar\'e descriptor $P_{k}\left(\lambda\right)$ of an individual pattern is the standard deviation of the intensity difference between rows separated by a fixed distance $k$, i.e.
\begin{equation}
    P_{k}\left(\lambda\right) = \textrm{std}\left(\sum_y I_{x,y}\left(\lambda\right) - \sum_{y} I_{x+k,y}\left(\lambda\right)\right).
\end{equation}
\noindent This Poincar\'e descriptor is otherwise known as the standard deviation of lag-$k$ differences, related to the standard deviation of successive differences \cite{vonNeumann41}, which is widely used in areas such as heart rate variability analysis \cite{Stein94,Brennan01}. It was recently demonstrated that this metric can probe short-range correlations in the spatial and temporal structure of speckle \cite{Majumdar17,Majumdar18}. In order to track grain-scale changes of speckle patterns, we set the parameter $k$ to the size of a typical speckle in the image, determined using the methods outlined in \cite{Majumdar17}. The variation with wavelength must be determined empirically, via a training phase which extracts a scaling law between $P_{k}$ and $\lambda$.

To demonstrate the method, we first simulate the speckle patterns produced for each of a range of wavelengths incident upon a multiply-scattering medium, using the model previously presented in \cite{Metzger17}. The simulation assumes a linearly-polarised and monochromatic incident Gaussian beam of a particular power, wavelength and fixed spot size. To generate speckle, the optical field is propagated through a path of multiple equally-spaced random phase plates using paraxial wave theory and the split-step beam propagation method \cite{Metzger06}. Each phase retarder is a matrix of randomly-assigned, spatially slowly-varying phase retardations, with a small refractive index difference to the air ($\Delta n=0.001$) ensuring the prominent effect is to randomly scatter light in the forward direction. The number and separation of the retarders is set to mimic a \SI{38.1}{\milli\meter}-diameter, spectralon-coated integrating sphere, with the number given by the average number of reflections within the sphere (calculated from the sphere multiplier \cite{Parretta13} to be 20) and the separation set to the sphere diameter. The resulting intensity distributions are discretized to integers between 0 and 255, and a spatial subset is sampled onto a $256\times256$ pixelated grid, to mimic the limited dynamic range and spatial extent of typical cameras.

\begin{figure}[!t]
\centering
\includegraphics[width=1\linewidth,trim={0 0.1cm 0 0cm}]{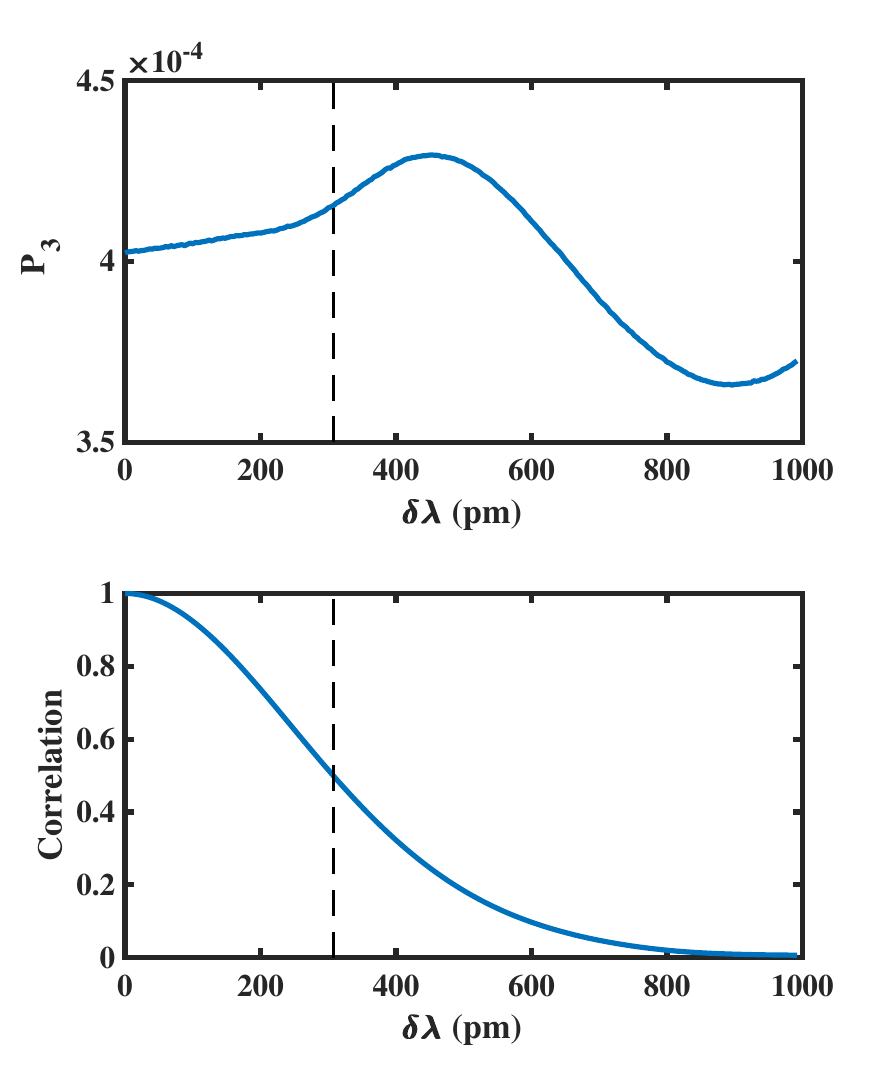}
\caption{The Poincar\'e descriptor $\textrm{P}_{k}$ (top) and speckle correlation (bottom) variation with wavelength, where $\delta\lambda=0$ corresponds to a wavelength of \SI{780}{\nano\meter}. The speckle correlation limit, above which the transmission matrix method is a reliable tool for speckle-based wavelength measurement, is indicated by the vertical dashed line. $\textrm{P}_{k}$ is monotonic over a wavelength range comparable to this speckle correlation limit, and therefore provides a reliable probe of wavelength below this limit. \label{FIGrange}}
\end{figure}

Speckle patterns for wavelengths between \SI{780}{\nano\meter} and \SI{781}{\nano\meter} (in \SI{4}{\pico\meter} steps) are generated for Poincar\'e analysis. The exact form of $P_{k}\left(\lambda\right)$ depends on the microscopic detail of the scattering structure, the image size and the value of $k$. Therefore, for a particular experimental apparatus to function as a wavemeter, $P_{k}\left(\lambda\right)$ must first be determined in a training phase.
An example of the dependency of $P_3$ on wavelength is shown in Figure~\ref{FIGrange}(a). In this example, $P_k$ varies monotonically with wavelength over a range $\sim\SI{430}{\pico\meter}$. Therefore, for wavelength changes smaller than this bandwidth, the Poincar\'e descriptor is a valid metric for extracting the wavelength. The bandwidth remains approximately constant for changes of image acquisition parameters and $k$, provided that $k$ is comparable to and above the speckle grain size. However, the upper and lower limits of the monotonic range can be tuned by judicious choice of $k$ and image acquisition parameters. This flexibility can be exploited during a training phase, in order to centre the range of operation around a wavelength of interest. 

The finite bandwidth is related to the speckle correlation length. The speckle patterns produced by two distinct wavelengths are similar for small wavelength separations, which gives rise to the continuous variation of $P_{k} \left(\lambda\right)$. However, the speckle patterns are highly dissimilar for large wavelength separations. The scale over which the speckle decorrelates is known as the speckle correlation limit, and gives a lower-bound to the resolution of the transmission matrix method for wavelength characterisation \cite{Cao17}. This is quantified as the wavelength separation for which the speckle patterns have a correlation coefficient of $50\%$. Figure~\ref{FIGrange}(b) shows the Pearson correlation coefficient between the speckle pattern produced with each wavelength in our test set, and the speckle pattern at $\lambda=\SI{780.000}{\nano\meter}$. The speckle correlation falls to $50\%$ for a wavelength separation of $\SI{307}{\pico\meter}$. This highlights the complimentary nature of the Poincar\'e descriptor method and the transmission matrix method: to extract the wavelength, the former requires that the speckles patterns have a sufficiently high correlation coefficient, while the latter relies on the speckle patterns produced by two wavelengths being uncorrelated. Thus the Poincar\'e descriptor method is a natural partner for use in tandem with the transmission matrix method, to increase the dynamic range over which wavelength can be measured.

\section{Experimental demonstration of wavelength recovery using the Poincar\'e descriptor method}

\begin{figure}[!t]
\centering
\includegraphics[width=1\linewidth,trim={0 0.1cm 0 0cm}]{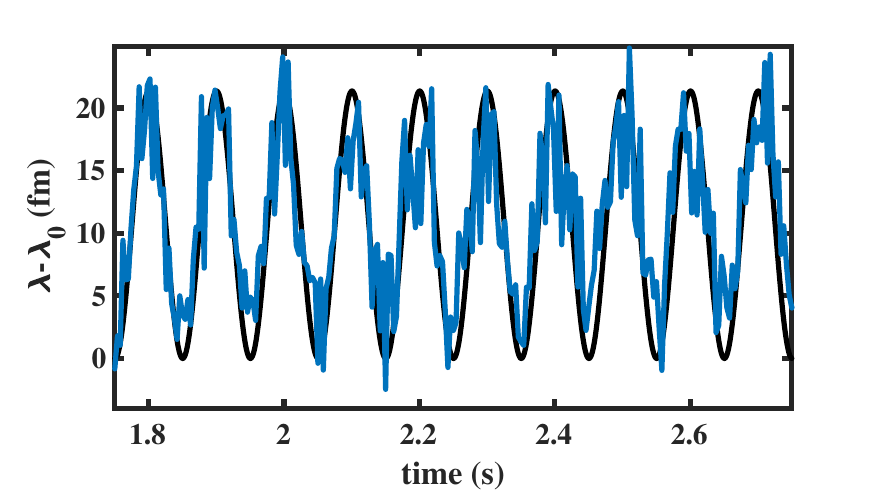}
\caption{In experiment, the Poincar\'e descriptor method can measure wavelength changes on the femtometer scale. A sinusoidal wavelength modulation with amplitude \SI{10.7}{\femto\meter} (black) is applied using an acousto-optic modulator, and the resulting changes in the speckle pattern are measured using the Poincar\'e descriptor $P_{3}$ (blue), achieving a signal-to-noise ratio of 3.1. \label{FIGexperiment}}
\end{figure}

We test the Poincar\'e method experimentally, using laser light from an external cavity diode laser (ECDL) (Laser diode: Toptica LD-0785-P220; Housing: Toptica DL-100) which is stabilised to the Rb-87 D2 line ($F=2 \rightarrow F=2\times3$ crossover) using saturated absorption spectroscopy, leading to a central wavelength of \SI{780.244}{\nano\meter}. In order to applied a controlled variation to the wavelength of the beam, we frequency-modulate the drive voltage of an acousto-optic modulator (AOM) (Crystal Tech 3110-120). The light is subsequently coupled into an angle-cleaved single-mode fibre (SMF) (ThorLabs P5-780PM-FC-10), eliminating any spatial variations in the beam. The SMF delivers \SI{1.5}{\milli\watt} via a physical-contact connector (ThorLabs SM1FCA) to a Spectralon-coated integrating sphere (Ocean Optics FOIS-1, diameter of \SI{38.1}{\milli\meter}, output aperture diameter of \SI{9.5}{\milli\meter}). The hollow sphere has a diffusely reflecting inner surface which causes the beam to disperse over time \cite{Boreman90}. After multiple reflections, light escapes through the sphere’s single output aperture and propagates in free space. The resulting speckle pattern is captured by a fast camera (Mikrotron EoSens 4CXP 4 megapixel CMOS camera) and sent to a computer to be analysed. The typical camera settings are a frame rate of 250\,fps, an image size of $256\times256$ pixels, and an exposure of \SI{8}{\milli\second}, and the distance between sphere and camera is chosen to be \SI{17}{\centi\meter} to ensure speckle grains are larger than the camera pixels. For Poincar\'e analysis, we use $k=3$ to match the size of a typical speckle grain from the sphere. Training is performed by recording the speckle patterns produced during 3 periods of a sawtooth waveform which modulates the wavelength over a range of \SI{40}{\femto\meter} at \SI{50}{\hertz} (i.e. the total training duration was \SI{60}{\milli\second}). From these, we extract $P_{3}\left(\lambda\right)$ for use in subsequent measurements. In Figure~\ref{FIGexperiment} we show the measurement of wavelength when a sinusoidal modulation is applied to the AOM with an amplitude of \SI{10.7}{\femto\meter} and a frequency of \SI{10}{\hertz}. The modulation is recovered with a signal-to-noise ratio of 3.1. 

\section{Comparison to Principal Component Analysis}

While the Poincar\'e descriptor method performs femtometer-resolved measurements, speckle wavemeters have previously been shown to resolve wavelength changes in the attometer-regime \cite{Bruce19}. Such high resolution required the use of Principal Component Analysis (PCA), a multivariate data analysis tool designed to identify a measurement basis from which maximal information can be extracted. 

PCA re-orients a high-dimensional dataset (i.e. the images of the speckle patterns produced as the wavelength varies) into an optimal basis from which the largest variations in the data are identified from comparatively few parameters (the Principal Components or PCs). As with Poincar\'e analysis, in order to extract unknown wavelengths, a training set of speckle patterns at known wavelengths must be collected. Each individual speckle pattern in the training set is flattened into a 1D array and independently normalised and the set is used to build up a matrix $A$, with each of the columns corresponding to an individual speckle pattern of a particular wavelength. The PCs of the training set are the eigenvectors $v$ forming the eigenbasis of the covariance matrix $M=A^{T}A$, where $A^T$ is the transpose of $A$, and these eigenvectors are sorted by decreasing magnitude of their corresponding eigenvalue. The calibration matrix $T_{cal}$ relating PCs to wavelength is given by $T_{cal}=Mv$. As the biggest variations in the training data are contained in the first few PCs, $T_{cal}$ can be down-sampled to a dimensionality of 
\begin{align}
    &\left(\textrm{the number of speckle images in the training set}\right) \nonumber \\ 
    & \times \left(\textrm{the desired number of PCs}\right).
\end{align}
\noindent Once this training procedure has been carried out, unknown wavelengths can be deduced by projecting their speckle patterns into the principal component space of the training set, 
\begin{equation}
\lambda = I_{x,y}\left(\lambda\right) T_{cal}. \label{EQN-PCA}
\end{equation}
\noindent The accuracy and range of the method depends on the density and number of images in the training set \cite{Mazilu14}, but is improved at the cost of computation speed.

To quantify the comparative speed of PCA and Poincar\'e analysis, we perform a wavelength measurement using a common set of speckle images using both approaches. For PCA this included solely calculating the PCs and for the Poincar\'e analysis this included finding $P_{k}$ for all frames. We performed Poincar\'e analysis for a range of values from $k=1$ to $k=200$, and found no dependence of the calculation time on $k$. We present data here for $k=1$. The analysis was performed on an early 2014 model MacBook Air (4Gb RAM, 1.4GHz processor), which has a comparable processor speed to available micro-controllers such as Raspberry Pi. We compare the speed for both of the necessary experiment phases for recording wavelength in a speckle wavemeter: the training phase and the measurement phase.

\begin{figure}[!t]
\centering
\includegraphics[width=1\linewidth,trim={0 0.1cm 0 0cm}]{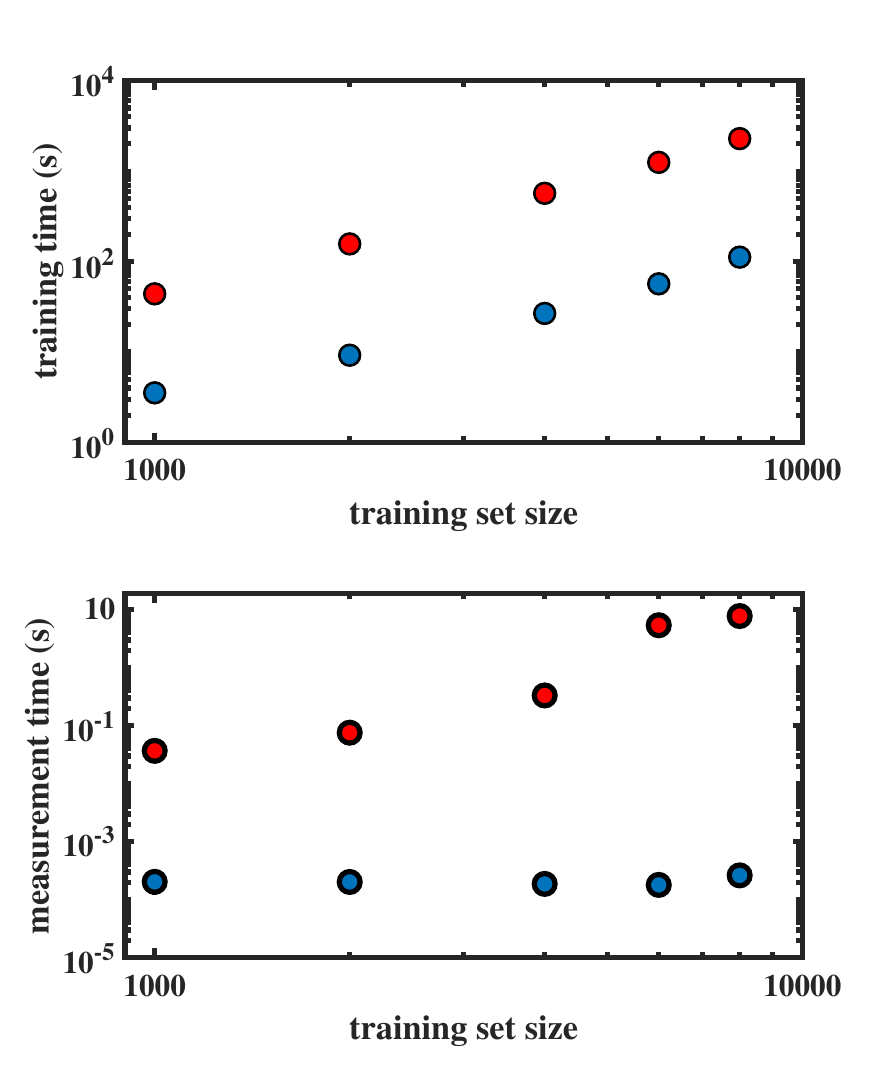}
\caption{Comparative measurement of the speed of PCA and of the Poincar\'e descriptor method. (a) The time required to train PCA (red) and Poincar\'e analysis with $k=1$ (blue) both exhibit power-law growth with the size of the training set. The training duration in Poincar\'e analysis is reduced by one order of magnitude compared with PCA. (b) After training has been performed, the time taken to perform an individual measurement is independent of training set size for Poincar\'e analysis, while the time for an individual wavelength measurement with PCA is at least two orders of magnitude longer and increases with training set size. \label{FIGspeed}}
\end{figure}

The time required for a training phase $t_t$ is shown in Figure~\ref{FIGspeed}(a), for training set size $s$ varying from 1000 images to 8000 images. The run time of both PCA and Poincar\'e analysis grow according to a power-law relationship, with
\begin{align}
t_t^{\textrm{PCA}} = \left(9\pm3\right)\times10^{-5}\times s^{\left(1.9\pm0.2\right)}, \nonumber \\
t_t^{P_{1}} = \left(6\pm3\right)\times10^{-5}\times s^{\left(1.6\pm0.1\right)},
\end{align}
\noindent where $t_t$ is the average time taken over three runs. The training duration in PCA is consistently larger by more than one order of magnitude.

The time taken to perform a wavelength measurement of a single speckle pattern using each technique was also found using the same MacBook Air. In the Poincar\'e analysis, the act of performing a measurement involves taking a particular speckle pattern, normalising it, flattening it and then finding the corresponding $P_{k}$ values. In PCA, the act of performing a measurement involves taking a particular speckle pattern and performing the projection into the PC space of the training set using equation~\ref{EQN-PCA}. As shown in Figure~\ref{FIGspeed}(b), for the Poincar\'e analysis the speed of measurement does not vary with training set size. However, for PCA the time to perform a measurement is significantly longer, and grows with training set size. Even for a modest training set comprising 1000 unique wavelengths, the measurement of an individual unknown wavelength takes \SI{37\pm1}{\milli\second} with PCA, more than two orders of magnitude longer than the \SI{230\pm40}{\micro\second}-duration measurement with Poincar\'e analysis. For this particular laptop, these correspond to maximum measurement rates of \SI{27}{\hertz} for PCA and \SI{4.3}{\kilo\hertz} for Poincar\'e analysis. 

In addition to the speed difference between PCA and Poincar\'e analysis achievable in a microprocessor, Poincar\'e analysis offers another potential advantage: the ability to carry out the analysis in an integrated circuit. To our knowledge, efficient computation of PCA in an integrated circuit has not been demonstrated without the use of microprocessors. However, the Poincar\'e descriptor $P_{k}$ is arithmetically simpler, making it possible to perform using circuitry. $P_{k}$ can be rearranged \cite{Asimow10} as
\begin{equation}
    P_{k} = \sqrt{\frac{\textrm{var}\left(\tilde{I}_{x}\right)+\textrm{var}\left(\tilde{I}_{x+k}\right)-2 \textrm{cov}\left(\tilde{I}_{x},\tilde{I}_{x+k}\right)}{2}},
\end{equation}
\noindent where $\tilde{I}_{x}=\sum_y I_{x,y}$ and $\textrm{var}$ and $\textrm{cov}$ are the variance and covariance respectively, both of which have been demonstrated to be calculable with FPGA (Field Programmable Gate Array) boards \cite{Bailey13,Martelli11}. 

\section{Conclusion}

Poincar\'e analysis of the laser speckle produced by an integrating sphere offers a method of measuring wavelength at femtometer-level resolution over a range of \SI{400}{\pico\meter}. The method is complimentary to the widespread transmission matrix method, which provides accurate wavelength measurement over a wider range but with a resolution comparable to the range of the Poincar\'e analysis. Additionally, the method computes significantly faster on a microprocessor than the state-of-the-art Principal Component Analysis, opening up the possibility for high-speed (\SI{>4}{\kilo\hertz}) real-time measurements of wavelength from speckle which will allow faster feedback locking to the current of diode lasers. A further route to the miniaturisation of speckle wavemeters is the possibility of removing the need for a computer or microcontroller to analyse images altogether, and perform the analysis with FPGA boards. In future, the method may be extended to measure not only the wavelength of a monochromatic laser, but the spectrum of broadband light \cite{Cao17}, or other parameters of the laser beam such as polarization \cite{Kohlgraf10} or transverse mode composition \cite{Mazilu12,Mourka13}.

\section*{Acknowledgements}
We thank D Cassettari, Y Arita, M Facchin, P Rodr\'iguez-Sevilla and P Wijesinghe for technical assistance and useful discussions, and acknowledge funding from the Leverhulme Trust RPG-2017-197. 

\bibliography{mybibfile}

\end{document}